\documentclass[preprint,12pt]{elsarticle}

\usepackage{amssymb}
\usepackage{amsthm}
\usepackage{graphics}
\usepackage{float}
\usepackage{tcolorbox}
\journal{Information and Software Technology}

\begin{document}

\begin{frontmatter}


\title{Work-from-home and its implication for project management, resilience and innovation - a global survey on software companies}

\author[USN]{Anh Nguyen-Duc}
\ead{anh.nguyen.duc@usn.no}
\author[Unibz]{Dron Khanna}
\ead{dron.khanna@unibz.it}
\author[QUB]{Des Greer}
\ead{des.greer@qub.ac.uk}
\author[Unibz]{Xiaofeng Wang}
\ead{xiaofeng.wang@unibz.it}
\author[UFSCAR]{Luciana Martinez Zaina}
\ead{lzaina@ufscar.br}
\author[ORT]{Gerardo Matturro}
\ead{matturro@fi365.ort.edu.uy}
\author[Unibz]{Jorge Melegati}
\ead{jorge@jmelegati.com}
\author[Unibz]{Eduardo Guerra}
\ead{eduardo.guerra@unibz.it}
\author[Oslomet]{Giang Huong Le}
\ead{GiangHuong.Le@oslomet.no}
\author[HEL]{Petri Kettunen}
\ead{petri.kettunen@helsinki.fi}
\author[LUT]{Sami Hyrynsalmi}
\ead{Sami.Hyrynsalmi@lut.fi}
\author[SDU]{Henry Edison}
\ead{hedis@mmmi.sdu.dk}
\author[PUCRS]{Afonso Sales}
\ead{afonso.sales@pucrs.br}
\author[BAL]{Didzis Rutitis}
\ead{didzisr@gmail.com}
\author[HEL]{Kai-Kristian Kemell}
\ead{kai-kristian.kemell@helsinki.fi}
\author[UMBC]{Abdullah Aldaeej}
\ead{aldaeej1@umbc.edu}
\author[JYU]{Tommi Mikkonen}
\ead{tommi.j.mikkonen@jyu.fi}
\author[UPM]{Juan Garbajosa}
\ead{juan.garbajosa@upm.es}
\author[JYU]{Pekka Abrahamsson}
\ead{pekka.abrahamsson@jyu.fi}
\address[USN]{Department of Business and IT, University of South Eastern Norway}
\address[Unibz]{Faculty of Computer Science, Free University of Bozen-Bolzano, Italy}
\address[QUB]{Queen\'s University Belfast, Northern Ireland}
\address[UFSCAR]{Universidade Federal de São Carlos, São Carlos, Brazil}
\address[ORT]{Universidad ORT Uruguay, Montevideo, Uruguay}
\address[Oslomet]{Faculty of Social Sciences, Oslo Metropolitan University, Norway}
\address[HEL]{University of Helsinki, Finland}
\address[SDU]{University of Southern Denmark, Odense, Denmark}
\address[LUT]{University in Lappeenranta, Finland}
\address[PUCRS]{Pontifícia Universidade Católica do Rio Grande do Sul, Brazil}
\address[UMBC]{University of Maryland Baltimore County, USA}
\address[BAL]{BA School of Business and Finance, Latvia}
\address[UPM]{Universidad Politécnica de Madrid, Spain}
\address[JYU]{University of Jyväskylä, Finland}
%\address[USN]{Department of Business and IT, University of South-Eastern Norway \\ Lærerskoleveien 40, 3679 Notodden, Norway}
%\address[UIA]{Department of Information Systems, University of Agder \\ Universitetsveien 25, 4630 Kristiansand, Norway}

\title{}


%\author{}

%\affiliation{organization={},%Department and Organization

\begin{abstract}
[Context] The COVID-19 pandemic has had a disruptive impact on how people work and collaborate across all global economic sectors, including software business. While remote working is not new for software engineers, forced Work-from-home situations come with both constraints, limitations and opportunities for individuals, software teams and software companies. As the "new normal" for working might be based on the current state of Work From Home (WFH), it is useful to understand what has happened and learn from that. [Objective] The goal of this study is to gain insights on how their WFH environment impacts software project and software companies. We are also interested in understanding if the impact differs between software startups and established companies. [Method] We conducted a global-scale, cross-sectional survey during spring and summer 2021. Our results are based on quantitative and qualitative analysis of 297 valid responses. [Results] We observed a mixed perception on the impact of WFH on software project management, resilience and innovation. Certain patterns on WFH, control and coordination mechanisms, and collaborative tools are observed globally. We find that team, agility and leadership are the three most important factors for achieving resilience during the pandemic. Although startups do not perceive the impact of WFH differently, there is a difference between engineers who work in a small team context and those who work in a large team context. [Conclusion] The result suggests a contingency approach in studying and improving WFH practices and environment in future software industry.
\end{abstract}



\begin{keyword}
Innovation \sep Work From Home(WFH) \sep Innovation Resilience \sep COVID-19 \sep Resilience



\end{keyword}

\end{frontmatter}


\section{Introduction}
Work from home (WFH) is increasingly being recognized as a popular work arrangement due to its many potential benefits for both companies and employees (e.g., increasing job satisfaction and retention of employees). WFH has been known in many different terms, such as remote work \cite{olson_remote_1983,olson_distance_2000}, virtual teams \cite{majchrzak_technology_2000,maznevski_bridging_2000}, and teleworking \cite{kurkland_advantages_1999,teo_empirical_1998,crossan_teleworking_1993}. Working remotely is not new in the software industry, and it is quite common to have team members from remote locations. However, with the recent COVID-19 pandemic, millions of people around the world were forced to adopt the WFH model, regardless of suitability to their business model, product nature, the current setting of teams and organizations \cite{neumann2021sars}. As estimated recently, there might be close to 40\% people working in the EU who began to telework full-time as a result of the pandemic \cite{eurofound_living_2021}.

Furthermore, many companies decided to switch to long-term remote work. In May 2020, Twitter's CEO informed their staff that they can work from home forever \footnote{https://www.forbes.com/sites/danabrownlee/2020/05/18/twitter-square-announce-work-from-home-forever-optionwhat-are-the-risks/}. Coinbase has become a “remote-first” company, allowing most staff who want to work remotely to do so indefinitely. Dropbox too will let all employees work from home permanently\footnote{https://www.businessinsider.com/what-spotify-twitter-goldman-sachs-said-about-long-term-remote-working-2021-3}and Amazon has stated that employees whose positions allow them to work from home can do so two days a week \footnote{https://www.flexjobs.com/blog/post/companies-switching-remote-work-long-term/}. While teleworking is not new in many organizational settings, recent studies in the context of COVID-19 agree that significant changes to the workplace or way of working will occur in post-pandemic times \cite{ford_tale_2021,smite_changes_2022,smite_forced_2021}. Organizations might face various configurations of WFH, from hybrid mode of office and online work to working from anywhere \cite{smite_forced_2021}.

On one hand, research about the relationship between working conditions and individual characteristics, i.e. productivity, well-being and work-life balance have been quite well explored in software project context \cite{nolan_work_2021,marinho_happier_2021,ralph_pandemic_2020,russo_predictors_2021,butler_challenges_2021,bao_how_2021}. These studies conceptualize constructs, such as well-beingness, fear, stress, distractions, communication and happiness and relate them to individual performance on different software engineering activities. It is shown that adopting WFH has a significant impact on individual well-being and is connected to their productivity and organizational support is needed to ensure better work-life balance for their employees \cite{molino_wellbeing_2020,galanti_work_2021}. 

On the other hand, research exploring WFH and organizational characteristics is relatively less explored in software engineering. The lockdown measures as a response to the pandemic threaten the existence of many innovative companies \cite{kuckertz_startups_2020}.
Particularly, for small businesses and software startups, it is essential to stay resilient during the pandemic period. Organizational resilience can be defined as the capability to handle challenging conditions in a way that secure an organization’s existence and prosperity \cite{vogus_organizational_2007}. In many cases, the implementation of WFH practices can enable organizations to continue operation and exist during the pandemic. However, some special working contexts cannot simply be shifted to WFH, rendering some organizations less resilient and potentially unable to operate \cite{bai_digital_2020}. Bai et al. found that WFH practices are critical for companies to maintain their operations. This effect might be heightened in scale for startup and small companies, due to their vulnerability to macro environmental changes. However, little is known about the resilience of startups during the shift to adopting WFH during the pandemic.
Inspired by this research gap, we are interested in exploring the working conditions of software engineers that was caused by the pandemic, and its implication to the operation and management of projects and companies. Besides managerial activities and resilience, we will look at how software companies maintain their innovation level during WFH circumstances. Tidd and Thuriaux-Alemán define innovation management practices as any structured administrative or technical help used to affect the effective implementation of the innovation process \cite{tidd_innovation_2016}. As collaboration and communication drive innovation behaviour, forced change by way of collaborating could affect companies' innovation management practices and other creative activities. In innovative sectors like the software industry, it is critical to understand how organizational innovation is affected by the new working situation.

Deriving from our multiple objectives, we will explore the following research questions, RQ1-RQ4 related to the effects of WFH on software companies. We are also interested in exploring the WFH situation in a software startup context. Consequently, the RQ5 is introduced to investigate if any special phenomenon can be observed in software startup companies.
\begin{itemize}
    \item RQ1 - How does the way of working in software projects change when WFH is adopted? 
    \item RQ2 - How is software project management impacted when WFH is adopted?
    \item RQ3 - How are  innovation activities in software companies  impacted by the WFH situation?
    \item RQ4 - How is resilience achieved by software companies when WFH is adopted? 
    \item RQ5 - Is there any difference between startups and established companies regarding the above impact? 
\end{itemize}

%We found a common pattern of work arrangement shift in our sample. Collocated work shift to work from home full-time or mostly full time. Distributed work across cities/ provinces increased. International collaboration remains the same. Company New control and coordination mechanism, ie. flexible meeting \& working time, new roles, etc. Many agree that innovation activities increase together with the new work arrangement and introduce fundamental changes to both process and business aspect of the companies in short future. Among investigated factors, we found that team and agility are among the most agreed factors that lead to organizational resilience.

\section{Related work}
This section briefly reviews existing work on the three key concepts: work-from-home, innovation management and resilience.
\subsection{Work from home}
WFH has been referred to using many different terms, such as teleworking, remote work and virtual team. Working from home is not a new phenomenon. Studies on the phenomenon has dated back to the 70s, when the motivation for WFH is either resource shortage and the will to reduce daily communication \cite{pratt_home_1984,crossan_teleworking_1993,majchrzak_technology_2000,bailey_review_2002,perez_perez_technology_2004}. WFH has been characterized as needing minimum physical requirements, individual control over work pace, defined deliverables, a need for concentration, and a relatively low need for communication \cite{olson_remote_1983}.

In general, WFH is often claimed to improve productivity \cite{cascio_managing_2000,davenport_two_1998} and teleworkers
consistently report increased perceived productivity \cite{duxbury_telework_1998,baruch_teleworking_2000}. WFH gives people more flexible working time and provides better work-life balance, save cost of central working place and might give better job satisfaction. While increasing productivity, WFH is found to be associated with greater levels of both work pressure and work-life conflict \cite{russell_impact_2009} because work intrudes into developers' home lives through working unpaid overtime, thinking about working hours, exhaustion and sleeplessness \cite{hyman_worklife_2003}. However, many organisations lack appropriate plans, supportive policies, resources or management practices for practising WFH.

\subsection{Work from home and software engineering}

Bao et al. conducted a quantitative analysis based on a dataset of developers’ daily activities from Baidu Inc \cite{bao_how_2021}. The authors found that WFH had both positive and negative impacts on developer productivity in terms of different metrics, e.g., the number of builds/commits/code reviews. Forsgren et al. studied open source projects in Github and showed that developer activity in terms of the number of pushes, pull requests, code reviews and commented issues remained similar or slightly increased compared to the pre-pandemic year \cite{forsgren_space_2021}.

Ralph et al. performed an extensive study of the pandemic impact on programming, including productivity in the early months of WFH \cite{ralph_pandemic_2020}. They concluded that perceived productivity has declined (admitting a marginal effect size) as a result of negatively affected well-being and that organizations need to accept that expecting normal productivity under  crisis circumstances is unrealistic. Ford et al. conducted a two-wave study on productivity in Microsoft \cite{ford_tale_2021}. Both surveys indicated that the productivity increased among some participants and stayed the same or decreased among the others, which led the authors to conclude that the productivity during the COVID-19 pandemic was dichotomous.

Russo et al. performed a two-wave longitudinal study with a diverse group of professionals, diving into the impact of over 50 psychological, social, situational, and physiological factors and their ability to predict the variance in well-being and productivity \cite{russo_predictors_2021}. The study concludes with a few associations between the studied factors and perceived productivity Oliveira et al. gathered data from two online surveys of Brazilian professionals \cite{oliveira_surveying_2020}. The authors found that perceived productivity in WFH when comparing with the office times has increased. Another important finding made regarding the changes in perceived productivity during the pandemic was that the number of positively affected respondents grew from 40\% in the first wave to 60\% in the second wave.

Smite et al. studied 13 surveys in the literature finding that on average perceived productivity had not changed significantly, there are developers who report being more productive, and developers are less productive when working from home \cite{smite_changes_2022}. Also, positive trends are found in longitudinal surveys, i.e., developers’ productivity in the later months of the pandemic show better results than those in the earlier months. Nolan et al. conducted a qualitative study on software engineering during the COVID-19 pandemic \cite{nolan_work_2021}. The authors showed that software companies will derive tangible benefits from supporting their employees during this uncertain time through ergonomic home offices, listening to their concerns, as well as encouraging breaks and hard stops to boost long term well-being and productivity. Machado et al. explored gendered experiences of eoftware engineers during the COVID-19 crisis \cite{machado_gendered_2021} and found that women face particular challenges during social isolation, as they lacked support with household and child care responsibilities.

\subsection{Software startups}
Berg et al. summarized a common definition on software startups as companies with innovation focus, lack of resources, working under uncertainty and time--pressure, highly reactive and rapidly evolving \cite{Berg_2018}. Steve Blank describes a startup as a temporary organization that aims to create high-tech innovative products without having a prior working history \cite{blank_steve_2010}. The author further highlights that in a startup context, the business and its product should be developed in parallel. Eric Ries defines a startup as a human institution that is designed to create a unique product or service under extreme uncertainty \cite{ries2011lean}. Rather than a formal company, a startup should be considered as a temporary organizational state, that seeks a validated and scalable business model \cite{Unterkalmsteiner_2015}. A company with a dozen employees can still be in a startup state to validate a business model or a market. 

Startups are found to be different from established companies in the strong presence of entrepreneurial personalities, behaviors, decision-making and leadership \cite{bygrave_theorizing_1992,khanna_mvps_2018,nguyen-duc_entrepreneurial_2021}. Software engineering literature also showed evidence on unique characteristics of product development in startup contexts. Giardino et al. revealed reasons for project failure in startups, in which many are not relevant to established companies \cite{Giardino_Fail_2014}. Nguyen-duc et al. conceptualized the co-development of product and business in startups as hunting and gathering activities \cite{Nguyen_HunterModel_2015}. Tripathi et al. found that entrepreneurs' background influence how startups' products are developed \cite{tripathi_startup_2019}.  Melegati et al. also showed evidence that startup founders have special kind of influences on requirement engineering activities \cite{melegati_model_2019}. Nguyen-duc et al. characterize the sense making processes in software startups, which is unique to the organizational states \cite{nguyen-duc_entrepreneurial_2021}.

\subsection{Innovation Activities in Software Companies}
%1. Innovation 1. What do we mean with "innovation", 2. What do we know about research on "product/ process innovation" in software companies, and in startup companies. 3. What1) Commitment to innovation, (2) Business investment on innovation,  (4) Intellectual Properties and innovation (5) business models

The process of innovation is defined as a way whereby novel and creative ideas are implemented to obtain business value for customers. It is the process of implementing external and internal ideas of the company and making them available to the market to advance the technology \cite{chesbrough2003open}. Innovation activities are not straightforward and involve several iteration cycles of many internal, external, and coupled activities \cite{gassmann2004towards}. The mechanism which should be quickly adapting to the changes happening with respect to the market situation \cite{munir2016open}. 
%%A well known and widespread concept of Lean Startup \cite{ries2011lean} practices the initial stage of prototype development and making it available to the early bird in the market for the feedback before the official release of software product or services \cite{fitzgerald2017continuous}. Innovation accounting which is one of the Lean Startup principles is a framework to estimate the progress of a startup .  Design thinking another known method of software engineering, outcomes innovative products and services. In other terms, the outside-the-box ideas. 
Design thinking is one example of an innovation process often seen in a startup context \cite{brown_design_2008,ries2011lean}. Software teams implementing the process of design thinking follow a creative and flexible way to create a solution to the problems elicited from the customers. In software projects, requirement elicitation is the process from which innovative ideas are built, as directly obtained from the customer \cite{bhowmik2016creativity}. These innovative ideas could be related to a software product, services, processes, and tools.  

\subsection{Resilience}
%Resilience 1. What do we mean with "resilience", 2. What do we know about research on "resilience" in software companies, and in startup companies. 3 concepts on factors like  Agility, Leadership, Financial Capital, Social Capital, Processes and Working Practices and Innovation

%resilience is the prevailing way of thinking in the area of sustainability studies, and the main contribution of resilience to sustainability is how to deal with uncertainties [Wen-Dong Lv].

In organizational contexts, resilience can be seen as a perseverance of systems and their ability to adapt to ongoing changes during the phase of disturbance \cite{holling1973resilience}. The major emphasis is on the persistence of the company, in other words to be able to survive in difficult situations \cite{lv2018innovation},\cite{moenkemeyer2012innovator}, \cite{lengnick2011developing}.  A turbulent period or critical situation can occur due to a technological change, an uncertain period in the company, complex regulatory problems, law or legal circumstances that affects the business  \cite{lv2018innovation}, and innovation project failure \cite{moenkemeyer2012innovator}. The six resilience constructs that are outlined by Moen et al. are: self-efficacy, outcome expectancy, optimism, hope, self-esteem and risk propensity \cite{moenkemeyer2012innovator}. The response to a critical situation can be either via adaptive fit or robust transformation. If the critical situation is predictable i.e. can be foreseen, the solution is an adaptive fit; otherwise a robust transformation is required \cite{lengnick2005adaptive}. 
Software startups in particular tend to
excel at adaptability and flexibility \cite{nguyen-duc_entrepreneurial_2021} and we should expect them to demonstrate that in response to the COVID-19
crisis. During the pandemic, it seems more appropriate for innovative startups to embrace iterative and flexible approaches such as effectual logic \cite{sarasvathy_causation_2001,nguyen-duc_entrepreneurial_2021}.

\section{Conceptual Framework}
Collin et al. reviewed and discussed different roles of theory in empirical work, and one of them is to provide focus and organization to the study \cite{collins_central_2018}. Maxwell et al. refer to a conceptual framework as ''an idea context of the study"  \cite{maxwell_qualitative_2012}. The conceptual framework should assist the researcher in refining goals, developing research questions, discerning methodological choices, identifying potential threats to validity, and demonstrating the relevance of the research. The primary source of the conceptual framework, from his perspective \cite{maxwell_qualitative_2012}, does not necessarily need to be an existing theory. Four primary sources are options from which to derive a conceptual framework: (1) knowledge based on experience, (2) existing theory, (3) exploratory research, and (4) “thought experiments” \cite{collins_central_2018}.

\begin{figure}[H]
\centering
\includegraphics[scale=0.16]{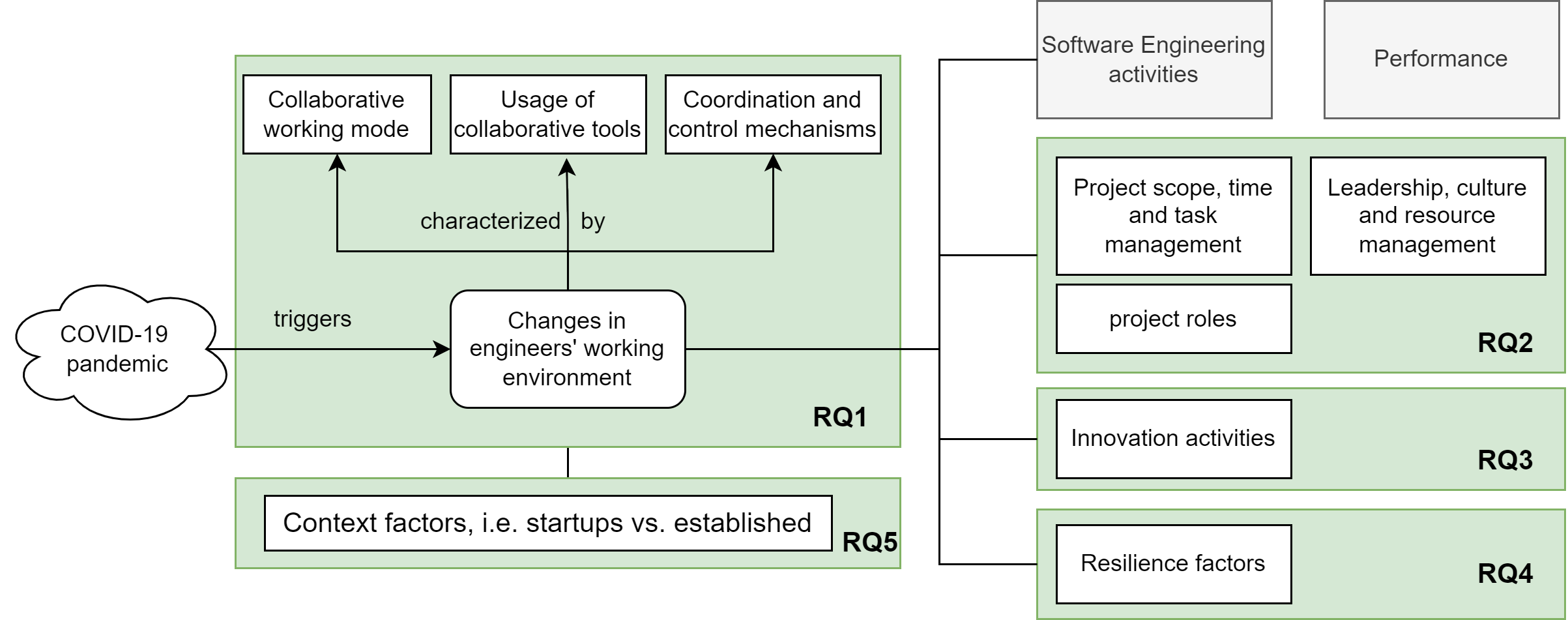}
\caption{\textbf{Conceptual framework}}
\label{fig:conceptualframework}
\end{figure}

Figure \ref{fig:conceptualframework} describes five theoretical blocks in which we are interested in our research: (1) Engineers' working condition (RQ1), (2) project management (RQ2), (3) Innovation activities (RQ3), (4) Resilience factors (RQ4) and (5) Context factors (RQ5). We also collected inputs on Software Engineering activities and Performance metrics. However these elements are not included in the scope of this work.
WFH setting and coordination mechanisms; (2) software engineering; 3) Innovation and Resilience; and (4) Performance.

\textbf{Working condition of software engineers} depends on the policy and setting of their companies. There are three important aspects of a virtual working condition that have been explored in Global software development research \cite{herbsleb_global_2001,ebert_surviving_2001}, namely the distribution setting \cite{nguyen-duc_impact_2015}, the usage of tools \cite{lanubile_collaboration_2010}, and the coordination and control practices and mechanisms and approaches for control and management \cite{nguyen-duc_impact_2015}. The configuration of a virtual team is usually seen from the combination of the extent that a team spreads over geographical locations and timezones \cite{cummings_crossing_2009,nguyen-duc_impact_2015}. Collaboration tools are essential in a virtual team. Various common tools that are reportedly common in global software development \cite{lanubile_collaboration_2010} should be revisited in the context of forced virtual collaboration. We also take input from control and coordination practices that are previously reported in global teams \cite{smite_case_2005,kraut_coordination_1995,boden_coordination_2007,strode_coordination_2012}.

\textbf{Project Management}. Communication, collaboration and coordination, the major managerial activities in a software project, is expected to be significantly affected by the shift to a new working condition. It would also be reasonable to observe the impact of WFH on other aspects of project management, i.e planning and execution. COVID-19 added more challenges to project management, and more effort on managing overview of tasks, team members, maintaining team commitment is important for project managers in the new context \cite{ng_challenges_2021}. After initial studies on project management and COVID-19, we selected core parts of project management that might be relevant, which are time management, scope management, resource and competence, and leadership.

\textbf{Innovation activities}. As stated in Section 2.4, it is important to understand how the innovation activities are influenced during the shift to a new working condition. It is mentioned in a recent report from Mc Kinsey that innovation is important to unlock post-crisis growth to high-tech companies \cite{McKinsey_innovation_2020}. However, the report also showed that commitment to innovation in many companies has decreased as companies work through the COVID-19 crisis and focus on short-term issue. Dogan et al. highlighted the possible difference between improvisational change and planned change in work environment in order to support strategic innovation processes \cite{dogan_strategic_2017}. The important elements of a business, such as commitment to innovation, financial and resource investment to innovation activities, Intellectual Properties, and business models are surveyed in connection to the perceived change of working conditions in this study.

\textbf{Resilience factors}. As stated in Section 2.5, resilience is an interesting factor that can be significantly influenced by the change of working conditions due to the pandemic \cite{muller_covid-19_2020}. During the pandemic, it seems
more appropriate for innovative startups to embrace iterative and flexible
approaches such as effectual logic \cite{nguyen-duc_entrepreneurial_2021,sarasvathy_causation_2001}. To confirm our intuition or otherwise, we asked which factors are perceived as resilience factors according to participants. We collected known factors from organizational resilience literature, namely, team, agility, leadership, financial capitals, social capitals, process and working practices and tools and infrastructure as the options with which to answer the question.

Besides, these key aspects, we are also interested in various context factors, such as:

\begin{itemize}
 \item Company types: startups vs. established 
    \item Team size
    \item Geographical location (national and continental levels)
    \item Application domains
    \item Participants' roles
\end{itemize}

\section{Methodology}
The goal of the survey was to gain insights into how the way of working has changed into WFH and how it impacts software engineering, project management, innovation and resilience.
\begin{figure}[H]
\centering
\includegraphics[scale=0.20]{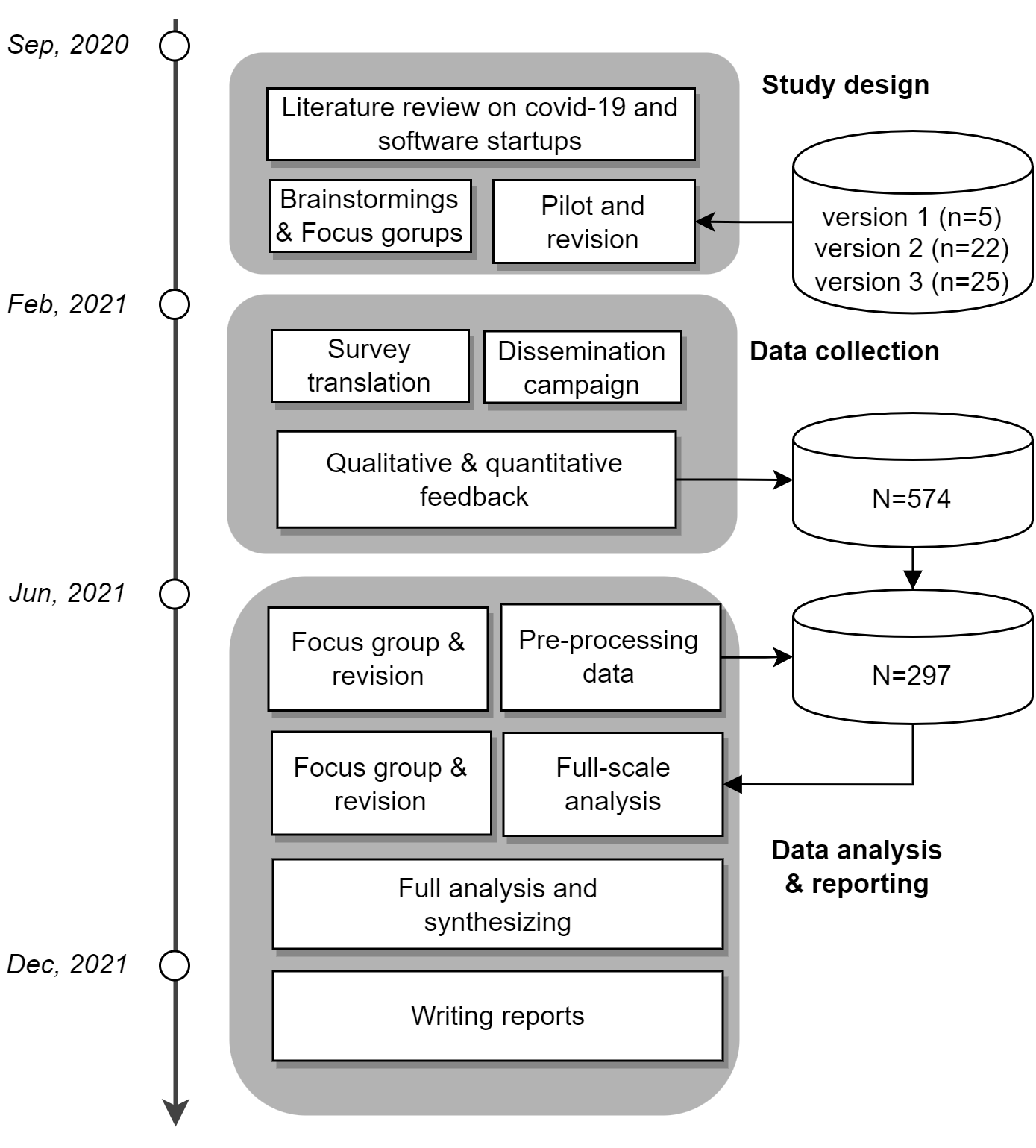}
\caption{\textbf{Research process}}
\label{fig:researchprocess}
\end{figure}

\subsection{Survey Design}
We designed a structured cross-sectional only survey to answer our questions.
The survey design process started in October 2020 and ended in Mar 2021. The initial foundation for the studies was a set of few studies about COVID-19 and startups or software engineering. It is noted that at the time the study was designed, there are not many papers found in this topic. The literature was updated gradually during the analysis and report writing. The study design was influenced by a large group of researchers. Typically, a brainstorming section for study design involved between 15 to 25 researchers from Brazil, Norway, Italy, Finland, Sweden, UK, Portugal, Germany, Australia, Canada, China and Vietnam. Many of these people are co-authors of this work. We conducted bi-monthly meetup via video-conference, either for brainstorming or focus-group to design the survey, discuss and work on the project. The result of this design process is three major versions of the survey (details on piloting and validating these versions are shown in Table \ref{tab:version}).

\begin{table}[H]
    \centering
    {\small
    \begin{tabular}{|p{0.8cm}|p{5.1cm}|p{1.8cm}|p{2.5cm}|p{1.3cm}|}
    \hline
      \textbf{Ver.}& \textbf{Major Activities} & \textit{\# meetings}& \textit{sample size} & \textit{Tools} \\
       \hline
       \hline
1 & Testing and validating theoretical elements  & 5 meetings & 5 internal participants & Paper based\\
\hline
2 & Reducing the number of questions, adding opt-out options & 3 meetings & 25 pilot companies & Google form\\
\hline
3 & Adding more open-text questions, Removing one-person companies  & 2 meetings & 5 internal participants & Lime survey\\
\hline
4 & Revising questions, correcting inviting text,languages  & 2 meetings & 517 responses & Lime survey\\
 \hline 
    \end{tabular}}
    \caption{\textbf{Major versions of the survey}}
    \label{tab:version}
\end{table}

The study’s target population is software developing companies anywhere in the world who switched from working in an office to working from home because of COVID-19. Stakeholders who had been working remotely before the pandemic are also important, but this study is about the switch, and the questions are designed for people who switched from working on-site to WFH.
The unit of analysis in this study is a software company, whether it is a startup or an established company. We implemented several approaches to making sure each participant can validly represent their companies. In principle, the questionnaire was open to a wide range of software development stakeholders, from business analysts, designers, software developers, testers, Scrum masters to startup-specific roles, such as CEOs or CTOs.

\subsection{Instruments}
The overall instrument used in this research constitutes in total 45 questions. The survey includes sections designed to (1) understand the current working conditions of the participants, (2) the contextual background of the participants, including the company and individual characteristics, (3) the perception of the participants about the impact of COVID-19 on software engineering activities in their team and companies and (4) the perception of the participants about the impact of COVID-19 on their companies' innovation and resilience and (5) perceived performance. 

We used both yes-no questions, multiple-choice questions (application domains, digital tools, team size and etc.), and five-point Likert scale multiple-choice questions. Some questions, for example, regarding digital tools participants are using, employed a free text option, e.g., “tool names” or “other” so that the respondents can specify their choice better. We also use several open text questions to get more details from the participants. In a Likert ordinal-scale question we have five standard choices: (1) Strongly disagree, (2) Disagree, (3) Neither disagree nor agree, (4) Agree Strongly and (5) Agree. Additionally, we added the sixth option (6) Not applicable so the participants can opt-out of the question. 

The interview questionnaire was developed in three months (as shown in Table 1). An initial draft (version 1) of the survey was first created based on the literature in software engineering and COVID-19. The first survey was created in Google forms. The second version of the survey was made after taking into account comments and adjustment from the whole author team. Some changes in later versions of the survey, for instance, re-coding scale labels of some questions, removal of few questions for better focus, adding team size/ company size value of 1 for filtering, adding open-text questions, and adding option "Not applicable" in questions. 

A pilot data collection was done in April 2021. We gathered responses from 25 companies to validate our constructs, scales and questions. We also asked for expert opinions from senior researchers in software engineering who conducted survey research before \cite{fernandez_naming_2017,ralph_pandemic_2020}. The final survey was made ready on May 2021 and available via the Lime survey tool. The major changes to versions of the survey is shown in Table 1. The details of the questions are available online \footnote{https://covidnse.limesurvey.net/561361?lang=en}.  

We have several questions helping us to determine whether a company is a startup or an established one. 

We also implemented  filtering questions to make sure only people who feel the impact of COVID-19 on their professional activities will continue filling in the survey: “Do you experience or observe an impact of COVID-19 to your work/ your company to any extent?”

The English version of the survey was translated into seven languages, which are Italian, Spanish, Portuguese, Norwegian, Arabic, Indonesian and Vietnamese. The translation was done by seven core members of the author team. The number of responses in each language is shown in Table \ref{tab:language}.

\begin{table}[]
\centering
\begin{tabular}{|p{4cm}|p{7cm}|} 
\hline 
\textbf{Survey language}    & \textbf{No. (digital or paper-based responses)}  \\ \hline
Italian            & 5                                      \\ \hline
Arabic             & 7                                      \\ \hline
Indonesian         & 11                                     \\ \hline
Chinese            & 12                                     \\ \hline
Spanish            & 13                                     \\ \hline
Vietnamese         & 42                                     \\ \hline
Portuguese         & 90                                     \\ \hline
English            & 394                                    \\ \hline
N=                 & 574    \\        \hline                       
\end{tabular}
\caption{\textbf{Responses in different languages}}
    \label{tab:language}
\end{table}

\subsection{Sampling Strategies}
Our sampling strategy is convenient and localized. We have tried several ways to purposefully gather the sample that can represent our target population. At the country level, we have had contact from 13+ countries, and we expect to get 10-20 responses per country via these personal channels. Conveniently, we invited participants through our professional and social networks. The invitation message was shared through co-authors’ social media, i.e. LinkedIn, Twitter, Reddit, Quora, and Facebook. We also published the call for participation in several academic communities, i.e. SEWorld, ICSOB, Software Startup Research Network, etc. We also explored our professional connections by asking co-authors to send invitation emails to those who they think likely able to participate. We capitalized on each author's local knowledge to reach more people in their jurisdiction. Rather than a single, global campaign, we used a collection of local campaigns. Each localization involved small changes in wording. In some cases, authors printed out the survey and disseminated the paper-based version instead. We also recruited participants from professional channels, such as Prolific \footnote{https://prolific.co/}.

\subsection{Data Collection}
\begin{figure}[H]
\centering
\includegraphics[scale=0.85]{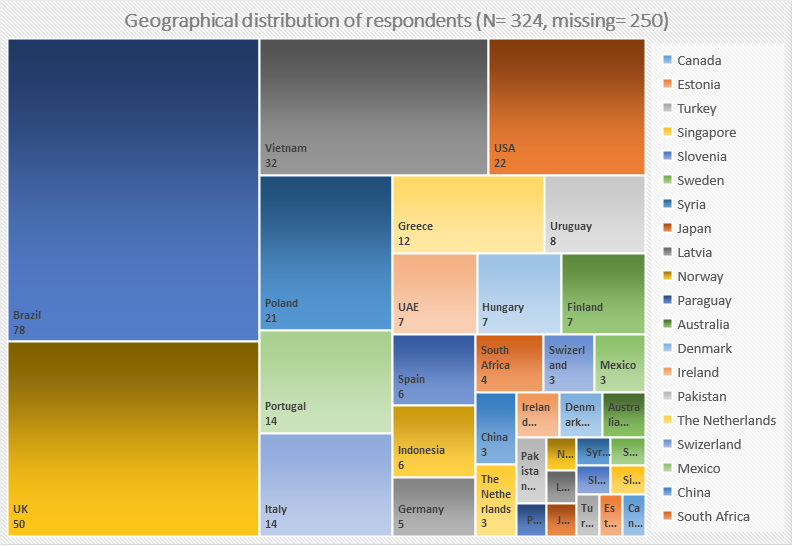}
\caption{\textbf{Graphical distribution of participants}}
\label{fig:geographicaldistribution}
\end{figure}

The data collection process started in March 2021 and ended in August 2021. In an invitation-based approach, we invited one respondent as a representative of the company. In a broad call for participation, we tried to control the sample representatives. Invitations led to a central survey tool, which can be configured to a suitable language version. The survey was spread in five online pages. As shown in Figure \ref{fig:researchprocess}, a total number of 574 responses were collected. 324 respondents were able to complete the survey (missing data rate is 43.55\%). The completion time ranges from 4 minutes to 155 minutes.
\begin{figure}[H]
\centering
\includegraphics[scale=0.8]{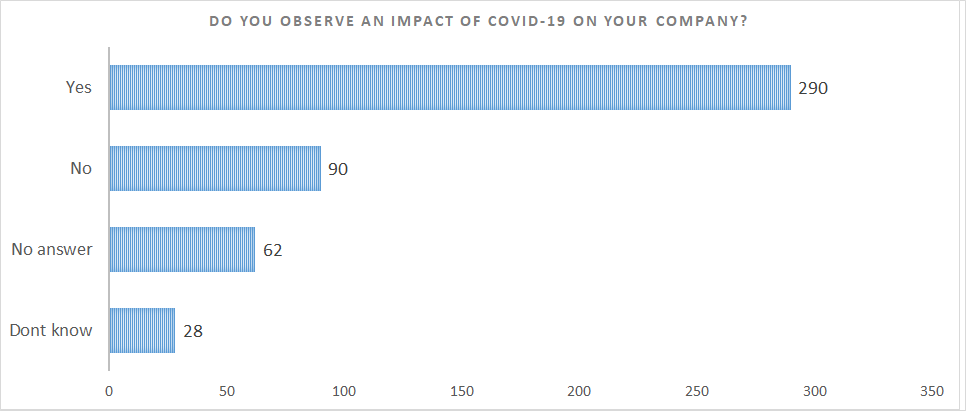}
\caption{\textbf{Awareness of COVID-19 impact on work}}
\label{fig:covidimpact}
\end{figure}
 A screening question was asked in the first place before other questions. If a respondent did not observe an impact of COVID-19 to their working environment in some ways, he or she is navigated to the end of the survey. In this way, the survey only gathers opinions from people who actually experienced the change to their working setting. As shown in Figure \ref{fig:covidimpact}, 290 people (50.5\%) answered 'Yes', while 90 people (15.7\%) answered 'No' to the questions. The rest had either no answer or answered 'do not know'.
As shown in Figure \ref{fig:geographicaldistribution}, the responses came from 35 countries, and dominated by respondents from Brazil, UK, Vietnam, USA and Poland.
After filtering irrelevant and invalid responses in data preprocessing step (Section 4.5.1), the number of valid responses for analysis is 297 responses.

\subsection{Data Analysis}

We received 574 total responses, and after data pre-processing steps, 297 valid responses remained.

\subsubsection{Data pre-processing}

The following steps were taken for data pre-processing:
\begin{enumerate}
    \item Remove responses with almost all empty answers,where the respondent apparently answered the filter question correctly, then skipped all other questions.
    \item Remove responses with almost all default answers, meaning that questions were not read properly
    \item Remove responses with outlined completion time. During pilot study, we know that the survey needs at least 3 minutes to be completed, and probably no more than 155 minutes. All answers that took less than 3 minutes or longer than 155 minutes are removed for quality reasons. 
    \item Remove responses from one-person companies (we had one question to identify company size)
    \item Filter responses that do not experience any changes due to the pandemic
    \item Move all free-text responses to a separate file for qualitative analysis
    \item Re-code raw data. We re-coded team size to categorize them into either small team versus large team context. We re-coded the company type (startup vs. established company), as described in Section 4.5.2.
    \item Re-code all quantitative answers to numeric value for quantitative analysis 
\end{enumerate}
 
\subsubsection{Identifying startup companies}
To allow comparative analysis, we need to have a way to identify startup companies. A challenge is that the term startup can be interpreted differently by people, especially when respondents are spread around the world. We asked different questions to identify the company's situations. Table \ref{tab:startupnumber} shows the result of the classification with 97 startups and 181 established companies. We have 19 participants who answered Unknown about the company type, and 277 who did not give their answers. 
\begin{table}[H]
\centering
\begin{tabular}{|l|l|}
\hline
\textbf{Company type} & \textbf{No. of responses}\\ \hline
Startup      & 97               \\ \hline
Established  & 181              \\ \hline
Unknown      & 19               \\ \hline
Missing      & 277              \\ \hline
N=           & 574            \\ \hline 
\end{tabular}
\caption{\textbf{Number of startups vs. established companies}}
\label{tab:startupnumber}
\end{table}

\begin{figure}[H]
\centering
\includegraphics[scale=0.8]{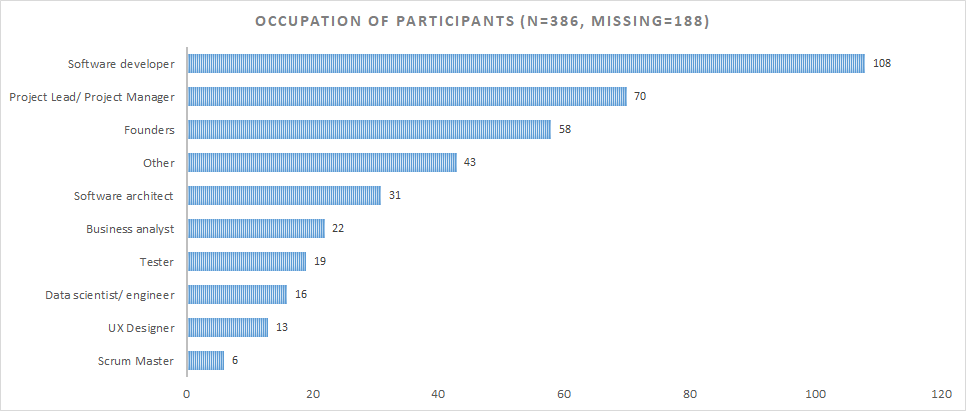}
\caption{\textbf{Occupational distribution among participants}}
\label{fig:roledistribution}
\end{figure}

Based on our definition (Section 2.3), we determined if a company is classified as a startup basing on the respondent's perception, the state of their main product and the company size. The classification rules are shown in Table \ref{tab:classrule}. 

\begin{table}[H]
\begin{tabular}{|p{2cm}|p{8.5cm}|p{2.5cm}|}
\hline
\textbf{Company types} & \textit{What is the last known state of   the product/service created in your company?}                                                    & \textbf{Status }     \\ \hline

Established                           & A product prototype is developed and has not yet been released   to market.                                                       & Startup     \\ \hline
Established                          & Not Applicable                                                                                                                    & Established \\ \hline
Established                          & Other                                                                                                                             & Established \\ \hline
Established                          & Product is rather stable, the focus is on gaining customer   base.                                                                & Established \\ \hline
Established                          & Product is stable, market size, share and growth rate are   established. Focus is set on launching new variations of the product. & Established \\ \hline
Established                         & Product was released to the market and is actively developed   further with customer input.                                       & Established \\ \hline
Startup                            & A product prototype is developed and has not yet been released   to market.                                                       & Startup     \\ \hline
Startup                             & Not Applicable                                                                                                                    & Startup     \\ \hline
Startup                             & Other                                                                                                                             & Startup     \\ \hline
Startup                             & Product is rather stable, the focus is on gaining customer   base.                                                                & Startup     \\ \hline
Startup                              & Product is stable, market size, share and growth rate are   established. Focus is set on launching new variations of the product. & Established \\ \hline
Startup                              & Product was released to the market and is actively developed   further with customer input.                                       & Startup     \\ \hline
Not sure                                & A product prototype is developed and has not yet been released   to market.                                                       & Startup     \\ \hline
Not sure                                & Not Applicable                                                                                                                    & Unknown     \\ \hline
Not sure                               & Other                                                                                                                             & Unknown     \\ \hline
Not sure                                & Product is rather stable, the focus is on gaining customer   base.                                                                & Established \\ \hline
Not sure  & Product is stable, market size, share and growth rate are   established. Focus is set on launching new variations of the product. & Established \\ \hline
Not sure                               & Product was released to the market and is actively developed   further with customer input.                                       & Startup \\ \hline    
\end{tabular}
\caption{\textbf{Classification rules for determining startups}}
\label{tab:classrule}
\end{table}

\subsubsection{Quantitative analysis}

For most of the questions, we conducted descriptive statistics to describe the distribution of respondents' answer. To answer RQ5, we need to compare respondents' answer among different categories. Because the dependent variable and independent variables are measured as categories at ordinal level, a Chi-square test will be used to examine the significant difference between groups. (https://stats.idre.ucla.edu/other/mult-pkg/whatstat/) 

\subsubsection{Qualitative analysis}
We applied a thematic analysis, which is commonly seen in empirical SE research \cite{Cruzes_Dyba_2011}. The objective of our thematic synthesis process was to add further details on the observations from quantitative analysis. Braun et al. suggest six steps for a thematic analysis: (1) familiarizing with data, (2) generating initial codes, (3) searching for themes, (4) reviewing themes, (5) defining and naming themes and (6) producing the report \cite{Braun_Clarke_2006}. 
\begin{itemize}
    \item Familiarising with data: the first and second authors of the paper read through all texts extracted from respondents to evaluate the quality level and possible themes. A total of 446 codes were obtained that were in English, Portuguese, Italian, Spanish, and Vietnamese languages. Text that was not relevant or did not give any useful content was excluded.
    \item Searching for themes: Instead of open coding that was done in SE research \cite{seaman_qualitative_1999,Wohlin_Aurum_2015}, we looked for texts that explain further why participants choose the answers for quantitative questions. For example, why one would choose Agility as the important factor for resilience, or why one would disagree on the increased role of IP during pandemic time. We identified 121 codes for Work-from-home, 84 codes for Project Management, 134 codes for Innovation, and 107 codes for Resilience mentioned in Figure \ref{fig:quali_fig}.
    \begin{figure}[H]
\centering
\includegraphics[width=9cm]{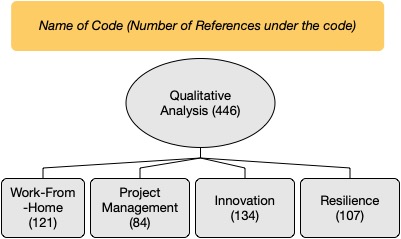}
\caption{\textbf{Qualitative analysis and reference codes}}
\label{fig:quali_fig}
\end{figure}

    \item Reviewing themes: We revised the codes and merged them into higher-level order codes. The codes that are not related to our quantitative findings will not be explored in this study.
    \item Producing the report: We created thematic maps for each theme, for instance, WFH (Figure \ref{fig:quali_wfh}), Project management (Figure \ref{fig:quali_pm}), Innovation (Figure \ref{fig:quali_inno}), Resilience (Figure \ref{fig:resilience}).
\end{itemize}

\section{Results}
This section presents answers to our RQs presented in Section 1 (Section 5.1 to RQ1, Section 5.2 to RQ2, Section 5.3 to RQ3, Section 5.4 to RQ4 and Section 5.5 to RQ5).

\subsection{RQ1 - How does the way of working in a software project change when WFH is adopted? }
Our data reveals the change to way of working in four different ways (1) shift in the collaborative working mode, (2) the usage of collaboration and communication tools, and (3) new coordination and control mechanisms. 

\begin{figure}[H]
\centering
\includegraphics[width=13cm]{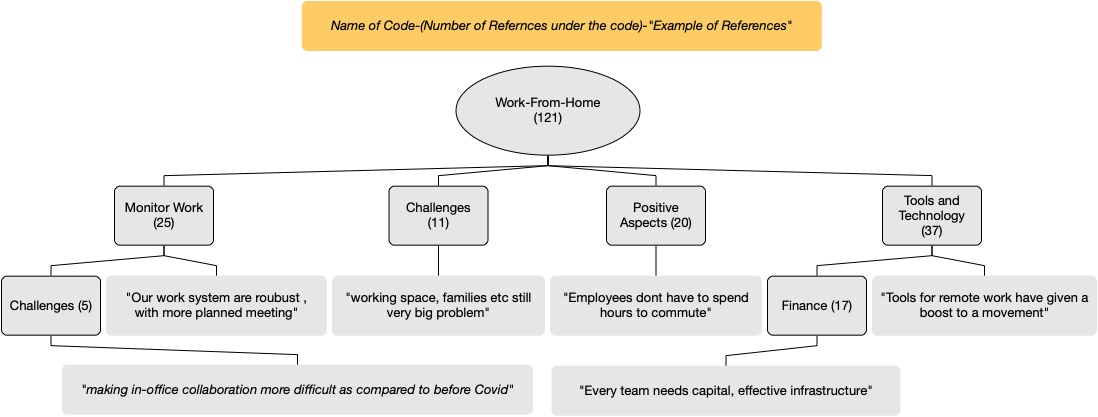}
\caption{\textbf{Thematic map of work-from-home and its reference codes}}
\label{fig:quali_wfh}
\end{figure}

\subsubsection{Shift in collaborative working mode}
\begin{table}[H]
    \centering
    {\small
    \begin{tabular}{|c|c|c|c|}
    \hline
     \textbf{Working} & \textbf{Virtual}  & \textbf{Before}  & \textbf{During} \\
           \textbf{modes}   & \textbf{setting}  &  \textbf{the pandemic}  & \textbf{the pandemic}  \\
           &&\textit{(in \% )}&\textit{(in \% )}\\
         \hline
         \hline
        Collocated &   Same office & 51.9 & 16.9\\
         \hline
           Same building & Different offices  & 24.7 & 13.8\\
         \hline
         \hline
        Closely with &Same  & & \\
        stakeholders &  city & 21.9 & 21.5 \\
         \hline
Closely with & Different & & \\
stakeholder &  cities & 22.3 & 33.5 \\
 \hline
 Closely with & Different & &\\
 stakeholders &  countries & 11.3 & 10.6 \\
\hline
\hline
Working  & Same &&\\
together & time zone & 30.3  & 37.1 \\
 \hline
 $<$ 8 hours & Different &&\\
differences  & time zones  & 10.2 & 14.5 \\
 \hline
 $\geq$ 8 hours  &  Different &&\\
differences & time zones  & 6.7 & 8.5 \\
       \hline  
    \end{tabular}}
    \caption{\textbf{Collaborative working mode across geographical locations and timezones}}
    \label{tab:developer}
\end{table}
There is a shift in working situation before and during the COVID-19 pandemic, as described in the Table \ref{tab:developer}. As expected, there is a significant reduction in the percentage of people who collocated in the same office (from 51.9\% to 16.9\%) and in the percentage of people who collaborate physically in the same building (from 24.7\% to 13.8\%). The virtual collaboration with stakeholders (i.e. customers, vendors, etc.) who are in the different cities or same time zones has increased significantly (from 22.3\% to 33.5\% and from 30.3\% to 37.1\% correspondingly). There is little change regarding whether collaboration occurred in the same city, collaboration across country boundaries or time zones.
\begin{table}[H]
    \centering
    {\small
    \begin{tabular}{|c|c|}
    \hline
      \textbf{Working from}& \textit{(in \% )} \\
       \hline
       \hline
   Home (fully) & 56.2\\
   \hline
   Home (mostly) & 18.02\\
   \hline
   Home (partially) \&  Office (partially) & 18.4\\
   \hline
   Office (mostly) & 6.36\\
   \hline
   Office (fully) & 8.8 \\
   \hline
   others& 0.7 \\
   \hline
   Not applicable & 2.09\\
   \hline
    \end{tabular}}
    \caption{\textbf{WFH porfolio of respondents at March 2021}}
    \label{tab:worklplace}
\end{table}

\begin{tcolorbox}
PO1: Collaboration across different cities and across different time-zones increased. Collaboration across the same city, and across different timezones tended to remain the same.
\end{tcolorbox}

\subsubsection{The usage of collaboration and communication tools}

\begin{table}[H]
    \centering
    {\small
    \begin{tabular}{|c|c|c|}
    \hline
      \textbf{Digital tools}& \textbf{Examples} & \textit{(in \% )} \\
       \hline
       \hline
       Video Conferencing & Teams, Zoom,  Meet& 87.27\\
       \hline
       Instant Messaging & Slack, Viber, Whatsapp & 79.8\\
       \hline 
       Cloud Storage & Private, One drive, & 57.95\\
       & Google drive, Dropbox & \\
   \hline
   Calendar Sharing & Outlook, Google Calendar & 55.12\\
   \hline
   Project Management Board & Team, Trello, Jira & 48.4 \\
   \hline
   File Sharing and & Github, Bitbucket & 45.22\\
 Version Control & & \\
   \hline
   Collaborative tools & Facebook & 24.02 \\
   for socializing & & \\
\hline
Collaborative tools  & Adobe Illustrator for UX design & 18.02\\
for specialized tasks & Overleaf for text editing & \\
\hline
Not applicable &  & 1.76\\
   \hline
    \end{tabular}}
    \caption{\textbf{Digital collaborative tools}}
    \label{tab:Digital}
\end{table}

 We also observed the WFH profile of respondents at the time the data was collected, as shown in Table \ref{tab:worklplace}. 56.2\% of the total number of respondents were fully WFH and only 8.8\% remained working in their physical offices. The respondents took to WFH in a positive way, evident from the 20 reference codes in Figure \ref{fig:quali_wfh} e.g. \textit{`` Employees do not have to spend hours to commute"} indicating that employees stayed home and utilized the commuting time in more effective manner. Other examples are: \textit{``Overall, it will be growth of our business"} and \textit{``Companies could reduce spending on office infrastructure"}. For some it already provided improved results e.g.  \textit{``It has sharpened our focus on delivery and generally been positive output"}.

While WFH employees used a variety of digital tools as presented in the Table \ref{tab:Digital}. This shows that 82.3\% of number of respondents frequently use video conferencing tools, for example Teams, Zoom, and Google Meet. It is probably a new way of working with increased usage of instant messaging software (79.8\% of total respondents), calendar sharing (55.1\%) and social media for professional socializing (24.2\%). It is also no surprise to see the adoption of cloud storage and specialized development tools in collaborative modes.

\begin{figure}[H]
\centering
\includegraphics[scale=0.7]{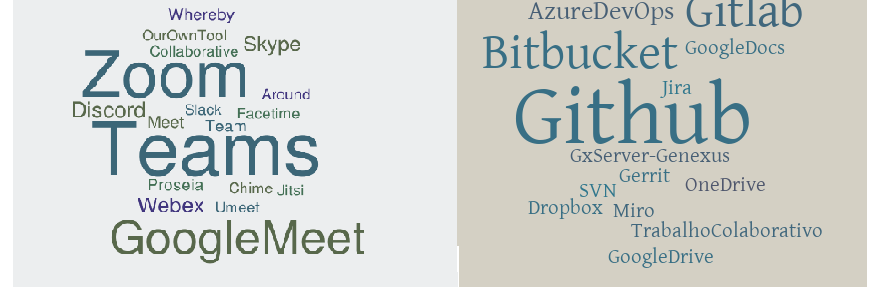}
\caption{\textbf{Most popular video conferencing tools and repository tools}}
\label{fig:wordcloud}
\end{figure}

Respondents shared how digital tools become more important for their companies:
\begin{itemize}
    \item \textit{``Tools for remote work have given a boost to a movement "} (Figure \ref{fig:quali_wfh})
    \item \textit{``We are embracing more collaborative tools for documentation; with so much communication online, people are more likely to use email/chat instead of calling meetings, allowing clients and devs to communicate more flexibly}" (Respondent389)
    \item \textit{``Increased use of chat and social media tools means developers facing issues can ask any colleagues regardless of their location for input/advice}" (Respondent371)
\end{itemize}
Figure \ref{fig:wordcloud} shows that Teams, Zoom and Google Meet are the most common tools for meeting when people working from distance. Other tools are also used, such as Skype, Whereby, Webex, Discord and Facetime. Regarding version control systems, almost all respondents reported the usage of a cloud-based solution. The most popular ones are Github, Bitbucket, Gitlab and Azure DevOps. 

Regarding calendar sharing options, the most common tools are Google Calendar and Microsoft 365/ Microsoft Outlook.

Regarding project management tools, Trello, Jira and Teams are the most common ones. Besides that, respondents also reported the usage of Asana, Asure DevOps, Monday, Redmine, Miro, and company internal tools.

Regarding socializing solutions, the popular social media tools are used, for example, Facebook, Linkedin, Twitter, Discord, Yammer, Sharepoint, and Instagram.

\begin{tcolorbox}
PO2: There are a wide range of tools being used to support collaboration, communication and project management during WFH.
\end{tcolorbox}
\subsubsection{New coordination and control mechanisms}

We gathered feedback from respondents on what new coordination and controls mechanisms are implemented as parts of their new working environment. It could also be an existing mechanism but has been adopted in a new frequency or format. Many respondents agreed about a significant change of coordination and control approaches in their companies:
\begin{itemize}
    \item \textit{``Members of my team are now, more than ever, required to document everything they do, in order to let other members of the team know what they are working on. That's positive. On the other side, brainstorming and interviewing people takes place remotely, making it harder to do effectively"} (Respondent380)
    \item \textit{``We had calls and meetings before COVID-19, a lot of informal meetups and syncs were done in office but now they have to be planned, scheduled, and organized. It slowed down some stuff that usually was very easy to solve"} (Respondent17)
    \item \textit{``Our work system are robust, with more planned meeting"} (Figure \ref{fig:quali_wfh})
\end{itemize}

\begin{table}[H]
    \centering
    \begin{tabular}{|c|c|}
    \hline
    \textbf{New mechanisms} &\textit{(in \% )} \\
    \hline
    \hline
       Changed and/or flexible times for working \& meeting & 55.1 \\   \hline
   Daily reports & 26.5 \\ \hline
   Changing the frequency of communication & 22.3\\
   via written documents e.g. emails, wikis, etc. & \\
   \hline 
Assigned new roles or responsibilities in the team & 21.9\\
\hline
Changing the frequency of retrospective meetings & 20.1\\
\hline
Adopting of online team coaching/ training & 19.8 \\
\hline
Daily clocking in/out & 18.7 \\
\hline
Using social media for team collaboration & 14.5 \\
\hline
Camera on while working & 6.3\\
\hline
Others & 6.7\\
\hline
Not Applicable &10.6\\
\hline
\end{tabular}
\caption{\textbf{Coordination and control mechanisms}}
\label{tab:controlcoor}
\end{table}

Table \ref{tab:controlcoor} shows the most adopted practices were to ease working and meeting times. More than half (55.1\%) of the respondents reported a change in working and meeting times implying more flexibility in managing their working time. Formal control mechanisms, for example regular reports (26.5\%), written documents (22.3\%), daily time registration (18.7\%) are in some cases adopted as a measures to the new working situation.

\begin{tcolorbox}
PO3: Various new coordination and control mechanisms are introduced and implemented for WFH, most of them based on formal procedures.
\end{tcolorbox}

\subsection{RQ2: How is software project management impacted when WFH is adopted}

\subsubsection{Scope, time and task awareness}

\begin{figure}[H]
\centering
\includegraphics[width=13cm]{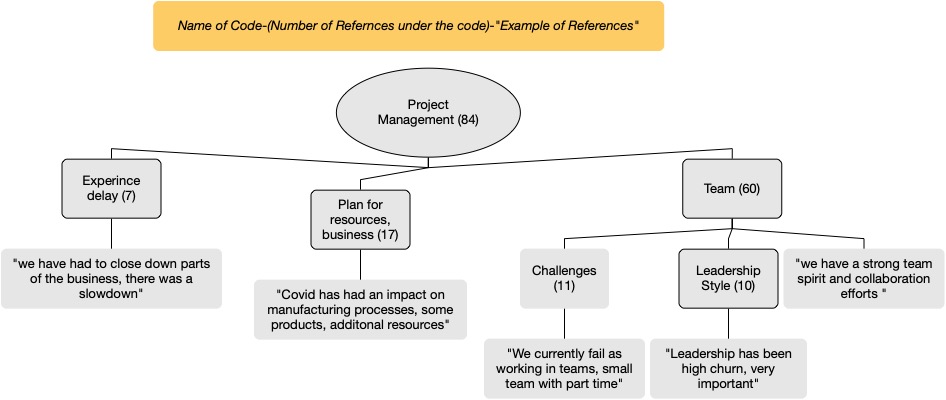}
\caption{\textbf{Thematic map of software project management and its reference codes}}
\label{fig:quali_pm}
\end{figure}

We asked whether respondents find it more difficult to have an overview of who does what in their current projects. 37.5\% of all respondents agreed while 28.7\% did not agree. 13.4\% answered neither agreed nor disagreed and 3.7\% did not see the question as applicable. We asked whether it became difficult to plan for resources, time, risk, and milestones. 41.2\% of all respondents agreed while 21.8\% did not agree. 18.1\% answered neither agree nor disagree and 2.8\% did not see the question as applicable. 
One respondent mentioned \textit{``Covid has had a n impact on manufacturing process, some products, additional resources"} (Figure \ref{fig:quali_pm}).

We asked whether the new working condition leads to a delay for their current projects. As shown in Table \ref{tab:softwaredelay}, 47.2\% of all respondents admitted some delays occurring in their current projects due to the shift. 13.4\% experienced significant delays and  3.2\% had their projects terminated. There was only 25\% who saw no impact on their project timeline. One respondent mentioned \textit{``We have to close down parts of the business, there was a slowdown"} (Figure \ref{fig:quali_pm}).

\begin{tcolorbox}
PO4: It is a mixed picture on how project planning and execution are perceived. However, the major tendency was to experience delays due to the working environment shift.
\end{tcolorbox}
\begin{table}[]
\centering

\begin{tabular}{|l|l|}
\hline
\textbf{Status of software projects}       & \textbf{Percentage} \\ \hline
Projects finished earlier than it would have been & 5.1\%               \\ \hline
Projects finished on time                   & 25\%                \\ \hline
Projects with some delays                  & 47.2\%              \\ \hline
Projects with significant delays           & 13.4\%              \\ \hline
Projects terminated                        & 3.2\%               \\ \hline
Not applicable                             & 6\%                 \\ \hline
\end{tabular}
\caption{\textbf{Status of software projects}}
\label{tab:softwaredelay}
\end{table}

\subsubsection{Leadership, culture and resource}

\begin{figure}[H]
\centering
\includegraphics[scale=0.6]{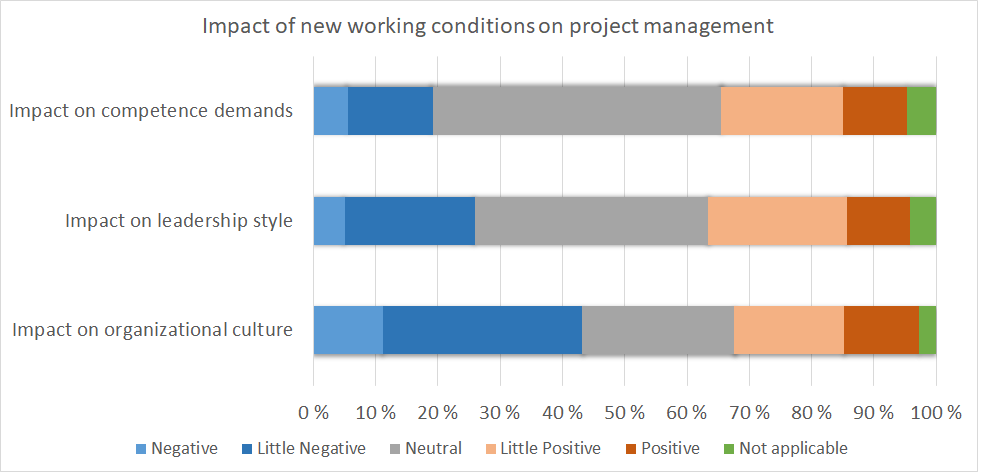}
\caption{\textbf{The perceived impact of WFH on competence demand, leadership style and organizational culture}}
\label{fig:leadershipcompetence}
\end{figure}

We also collected perceived responses about the impact of WFH on team-level aspects, such as team culture, leadership style and competence demands. Figure \ref{fig:leadershipcompetence} shows that regarding organizational culture, the respondents perceived little negative impact on team culture (31.9\% stated a little negative impact while 17.6\% answered a little positive). Regarding leadership styles and competence needs, there is no clear tendency observed.  A respondent staed that \textit{``Leadership has been high churn, very important"} (Figure \ref{fig:quali_pm}). About teamwork a project respondent mentioned \textit{``We have a strong team spirit and collaboration efforts"} (Figure \ref{fig:quali_pm}).

\begin{tcolorbox}
PO5: The perception of the shift to the new working environment on maintaining organizational culture tends more to negative impact.
\end{tcolorbox}

\subsubsection{Software Engineering roles}

\begin{figure}[H]
\centering
\includegraphics[scale=0.4]{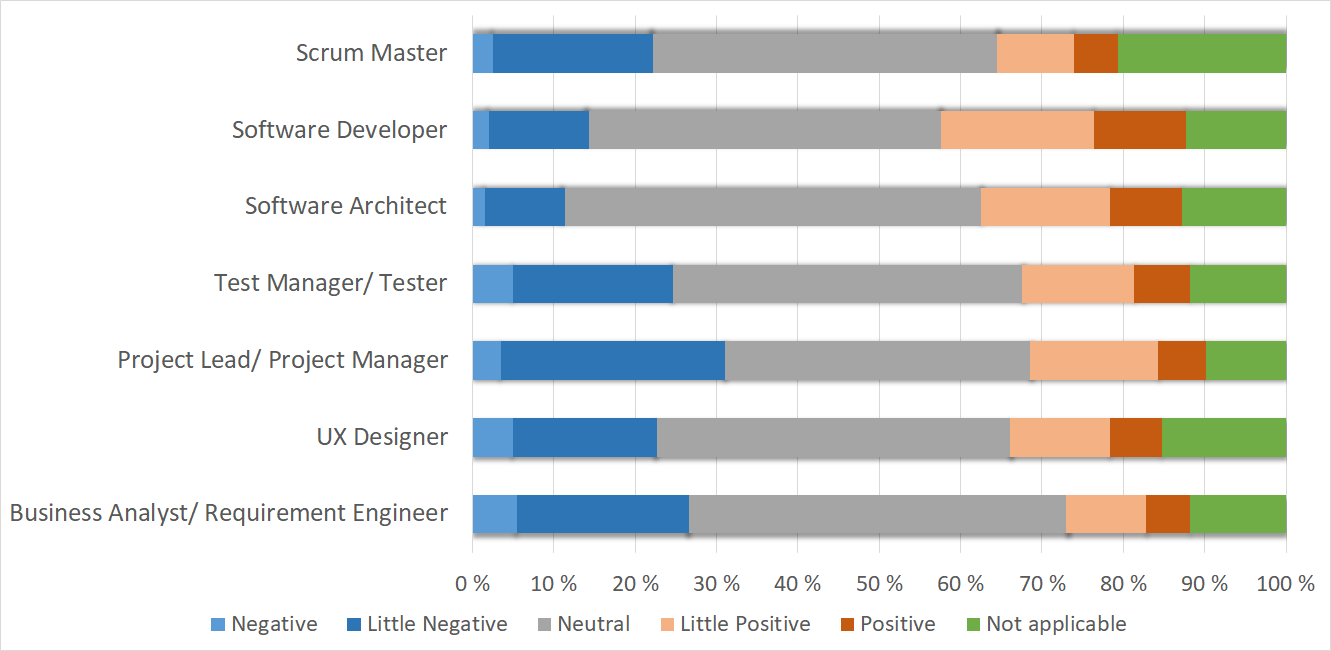}
\caption{\textbf{The impact of WFH on Software Engineering roles}}
\label{fig:serole}
\end{figure}

We asked according to the respondent's experience, how are the Software Engineering Roles affected due to the new working environment. The result is shown in Figure \ref{fig:serole}. For all SE roles that were named, there are a large number of answers (from 37.4\% to 51.2\%) saying no impact. Among these roles, the  negative impacts are mostly reported for project lead or project manager (31\%), business analyst (26.6\%) and tester (24.6\%). Among these roles, the positive impacts are mostly reported for software developer (30\%). 

Regarding manager's position, it seems that more managerial work and responsibility occur during the shift of team working conditions: \textit{About management work, it is harder to co-ordinate activity and to communicate with everyone.}(Respondent70)

\begin{tcolorbox}
PO6: Project managers and business analysts seem to be roles that are most impacted by the change to the new work environment. Software developers seem to be least impacted.
\end{tcolorbox}

Business analysts reported both negative and positive about their impact of new working conditions on their job, both within their team and with customers/ users:
\begin{itemize}
    \item \textit{``The good thing is that customers now arrive at a better distributed frequency throughout the day }" (Respondent46)
    \item \textit{``People are less formal and easier to access, to listen. During the meeting because we see each other on cameras, BO are less distracted with their phones and more focused in what we discussing. Validations are faster and people more open, more real, more productive in general. There is a learning curve, but we are learning different ways to share the information better.}" (Respondent318)
    \item \textit{``The pandemic has taken its toll on us so everything is a little more difficult than it used to be, including communication, the feel of being a team, interviews about a place in our company.}" (Respondent82)
    \item \textit{``For customers, soliciting feedback has become laborious when done remotely, so mainly a time issue}" (Respondent60)
    \item \textit{``Not being physically in the same room impinges upon the sharing of ideas and suggestions, it feels less natural and so it seems slightly stifled as a creative working environment}" (Respondent260)
    \item \textit{``Hard to get hold of stakeholders/customers since they are not in their office}" (Respondent140)
    \item \textit{``We think that making an interview or understand our clients problems by using programs as Zoom instead of meeting in person may reduce the effectiveness of the reunions}" (Respondent247)
\end{itemize}

\subsection{RQ3 - How are innovation activities in software companies impacted by the WFH situation?}
\begin{figure}[H]
\centering
\includegraphics[width=13cm]{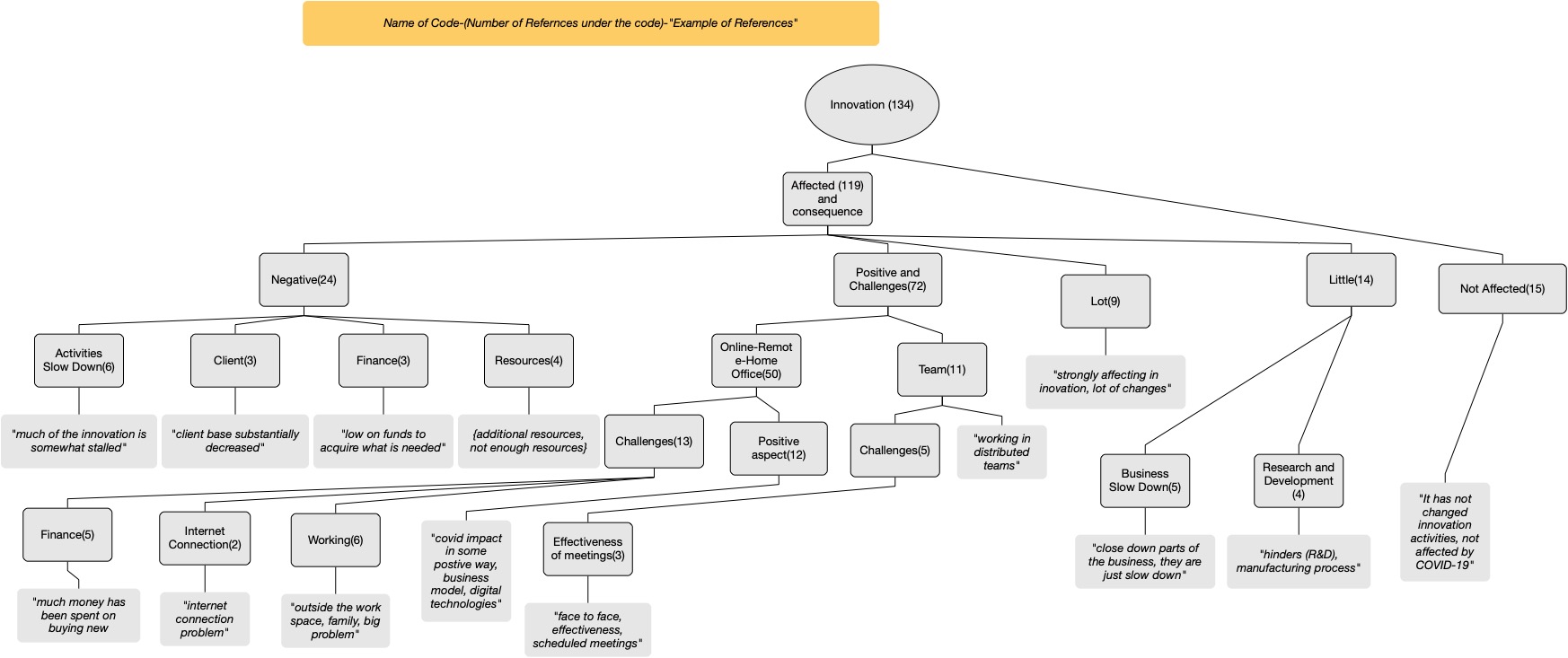}
\caption{\textbf{Thematic map of innovation and its reference codes}}
\label{fig:quali_inno}
\end{figure}

We found 134 coding references that refer to innovation activities in software companies.
119 coding references point out innovation was \textit{``affected"} due to COVID and 15 references mentioned they were \textit{``not affected"}with one respondent replying \textit{"It has not changed innovation activities"} (Figure \ref{fig:quali_inno}). Among affected respondents,  14 said \textit{``little"}, 9 were affected a \textit{``lot"} while 24 were affected in a \textit{``negative"} way  and 72 affected in a  \textit{``positive"} way. An interesting finding reveals that various codes were similar under the category of affected a \textit {``lot and positive"} as replied by same respondents, while the codes were similar under the category of affected \textit{``little and negative"} for the same respondents. 

Figure \ref{fig:quali_inno} shows 24 \textit{``negative affected"} respondents and some specified the reasons. For example, for \textit{``Client, Finance, Resources, Activities slow down Team"} the respondents said:
\begin{itemize}
    \item \textit{``Due to COVID innovation process was slow but it did not completely stop"}
    \item \textit{``Much of the innovation is somewhat stalled"}
\end{itemize}
Another respondent reported that an additional cost was involved for resources and materials:
\begin{itemize}
    \item \textit{``There are not enough resources and new materials are not easy to find"}
\end{itemize}

Among 72 references who responded with \textit{``positive affected"} innovation  also named \textit{``challenges"} with working from home and team effectiveness. 50 aggregate coding references stated an \textit{``online-remote-home office"} preference which helped the companies to innovate but also included challenges like \textit{``internet connection, product delivery, finance, working spaces"}. 

11 aggregate references show that the \textit{``teams"} were a key ingredient for a company although the challenge was that the \textit{``effectiveness of the meetings"} was affected. As one respondents said \textit{``It really affects our scheduled meetings and the effectiveness of our meetings, and face to face meetings"}. 
15 aggregate coding reference outlined that Innovation was \textit {``not affected"}. Typical responses were: \textit{``It has not changed innovation activities",``not affected by COVID 19",``did not influence company much"}. This was in some cases attributed to the fact that respondents \textit{``were already working from home at the beginning of the start up"}. Even if the team members did not work at home in the beginning, they adapted to the work from home culture e.g. \textit{``Since we work from home COVID did not impact innovation"}. Further, some companies already had a digital working environment e.g. 
\begin{itemize}
    \item \textit{``My financial education company has always been digital, I believe that in some points the pandemic has not affected this area"}.
\end{itemize}

\begin{figure}[H]
\centering
\includegraphics[scale=0.9]{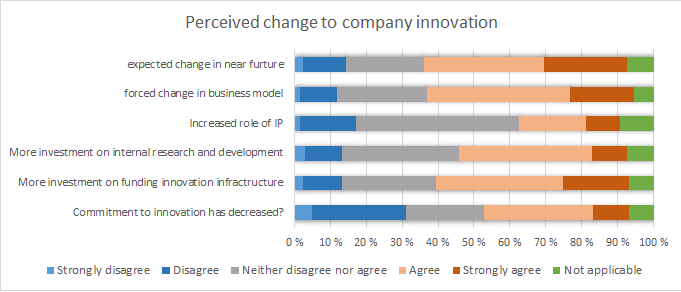}
\caption{\textbf{The perceived impacts of WFH on their companies' innovation activities}}
\label{fig:innovationchange}
\end{figure}

Five dimensions of innovation are investigated, namely commitment to innovation, investment on infrastructure, investment on internal R\&D, Intellectual Property (IP), and the changing business model. Figure \ref{fig:innovationchange} shows how respondents perceived the impact of working environment to these aspects.

Regarding commitment to innovation, 41.4\% of all respondents agree or strongly agree about a decrease in commitment to innovation, i.e. individual willingness to perform innovation activities within their teams. Examples are:
\begin{itemize}
    \item \textit{``Our client base substantially decreased over the last 12 months, subsequently much focus was on business profitability and continuation than innovation}" (Respondent56)
    \item \textit{``Maybe the greatest impact occurs in the low performance for the creation of new ideas. It has taken a lot for people to have new developments without being together and without generating synergies}" (Respondent101)
\end{itemize}
Only 31\% of respondents disagree or strongly disagree with this statement. e.g.:
\begin{itemize}
    \item \textit{``We have been developing new ways of communicating with customers. We have to innovate or we will go bust}" (Respondent140)
\end{itemize}

Regarding investment on infrastructure, 53.7\% of all respondents agree or strongly agree that there is an increase in companies' efforts and money to acquire new technologies and infrastructure. Only 13.3\% disagree or strongly disagree with this assertion. Some respondents said:
\begin{itemize}
    \item \textit{``The pandemic did not change the company's plans. However, it required to change some internal aspects of the organization and this led to investing in new technologies and infrastructures. even if imho, once the pandemic is over, everything will return to the way it was before}" (Respondent42)
    \item \textit{``By COVID-19 my company needs investments, for example for equipment for employees who need to work outside the office, and additional resources for masks and disinfectants}" (Respondent25)
    \item \textit{``So much money has been spent on buying new software and equipment for when we're working from home, so less is now available for funding elsewhere}" (Respondent15)
\end{itemize}

Regarding investment on internal R\&D activities, 46.8\% of all respondents agree or strongly agree that there has been an increase in companies' efforts and money for internal research and development. Only 13.6\% disagree or strongly disagree with this position.E.g.:
\begin{itemize}
    \item \textit{``The limitations regarding social gathering and face to face meetings, as a result of COVID-19, has caused the dynamics of the organization to change and adapt.  Some of this is because of the need for advancing technology and some is because of financial constraints}" (Respondent278)
\end{itemize}
There were neutral options provided when asking about the increase role of IP in their companies.
When asking about the new business models, 57.6\% of all respondents believed that COVID-19 appears as a factor forcing a change in their companies' business models. 58.7\% of respondents agreed or strongly agreed that COVID-19 will fundamentally change the way business is done in their companies' sectors over the next five years. Only 11.8\% and 14.3\%, respectively, take the opposite viewpoint.

\begin{tcolorbox}
PO7: It seems that at the individual level, the majority experienced an decrease in motivation for innovation. However, at the organizational level, many agreed that their companies invest more on innovation infrastructure.
\end{tcolorbox}

\subsection{RQ4 - How is resilience achieved by software companies?}
\begin{figure}[H]
\centering
\includegraphics[width=14cm]{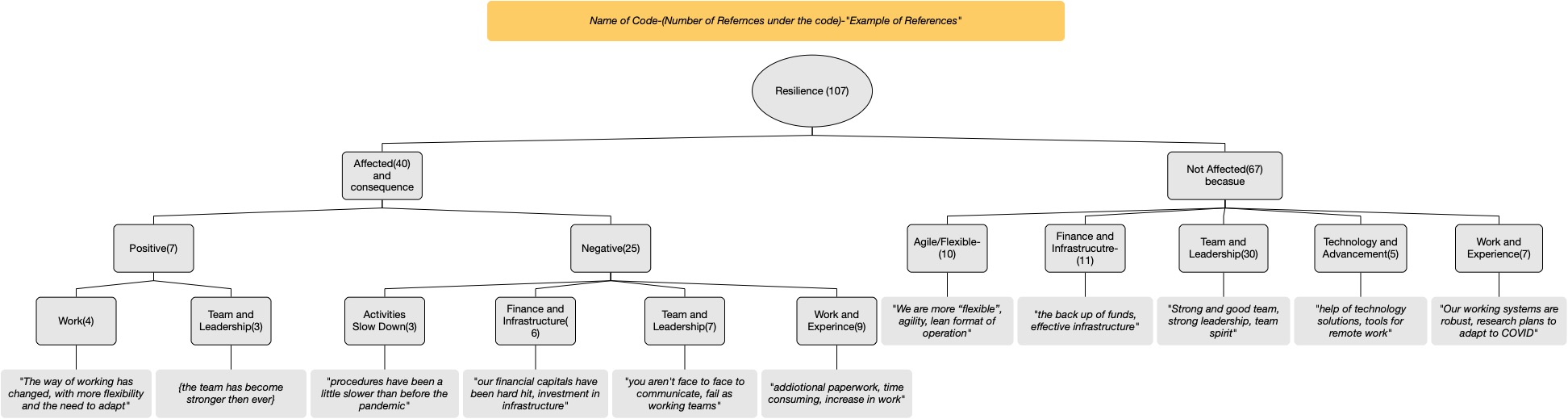}
\caption{\textbf{Thematic map of resilience and its reference codes}}
\label{fig:resilience}
\end{figure}

\begin{figure}[H]
\centering
\includegraphics[scale=0.55]{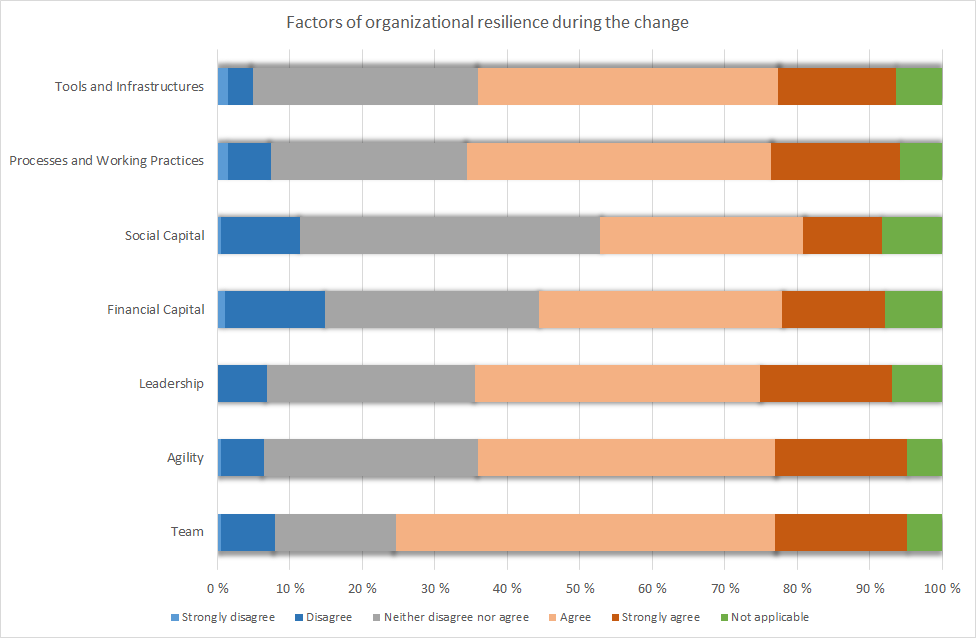}
\caption{\textbf{The most perceived resilience factors during WFH}}
\label{fig:resiliencefactor}
\end{figure}

We investigated seven factors from literature relating to organizational resilience, which are team, agility, leadership, financial capital, social capital, processes and working practices and tool and infrastructures. Figure \ref{fig:resiliencefactor} shows that overall, all of these factors are perceived helpful for organizational resilience. The most attributed resilience factors for software companies in our study are 'Team' (agreed by 71\%. E.g.: respondents)
\begin{itemize}
    \item \textit{``Luckily, we have a strong team spirit and collaboration efforts which help us to become more resilient}" (Respondent15)
    \item \textit{``We have a strong team spirit so that will get us through this difficult time. however, our financial capitals have been hard hit by the pandemic and navigating the uncertainty and down time over the past year}" (Respondent46)
    \item \textit{``We have very nice team we are working together and trying our bests to negate COVID-19 work problems}" (Respondent81)
    \item \textit{``I think people are more together than never, even tough they are not together, they are more friendly}" (Respondent85)
\end{itemize}
    
Agility (agreed by 59.1\% respondents) and Leadership (agreed by 57.6\% respondents) e.g. :
\begin{itemize}
    \item \textit{``We are more ``flexible". We're very small. So we don't have stuff to do 24/7. So we take our time, being ``flexible"}" (Respondent33)
\end{itemize}
Respondents highlighted the role of leaderhips e.g. :
\begin{itemize}
    \item \textit{``Team and Leadership properties have increased due to the management becoming more trusting and allow staff to work from home and more freedom}" (Respondent382)
    \item \textit{``We currently fail as working in teams at the moment but each member is doing their best to reach the goals of the organisation especially with the strong leadership in place}" (Respondent400)
    \item \textit{``We are all part of a big team and working together with good standards of leadership in place.  working practices have changed a little but the team has become stronger then ever to over come these difficult times}" (Respondent404)
    \item \textit{``Knowing how to make the right choices during the pandemic has made the company stronger than before. this above all thanks to a good team and good leaders who have been working in the sector for generations}" (Respondent42)
\end{itemize}

The factor that came out as least important was social capital (agreed by 38.9\% respondents).

The qualitative results explain further the observation from the survey answers. With 107 coding references, the Figure \ref{fig:resilience} show 40 \textit{``affected"} and 67 \textit{``Not affected"}. Hence in comparison less companies were affected by COVID-19. Among the affected, the 25 references show  \textit{``Negatively affected} and its consequences, whereas 7 references had \textit{``positively affected"} which outlines \textit{``Agile/Flexible, Team and Leadership, Technology and Advancement"} as the consequences to the companies. The \textit{``Negative affect"} outlines the \textit{``activities slow down, finance and infrastructure, team and leadership, technology and advancement, and work and experience} affects.  67 aggregate coding references outline \textit{``Not affected"}. This is due to the fact the companies to a certain extent had a good operation in relation to \textit{``Agile/Flexible, Finance and Infrastructure,Suppliers, Team and Leadership, Technology and Advancement, Work and Experience"}.

\begin{tcolorbox}
PO8: Team, Agility and Leadership are the most important factors for software companies to achieve organizational resilience. 
\end{tcolorbox}

\subsection{RQ5 - Is there any difference between startups and established companies regarding the above impact?}

\subsubsection{Project management}
Startup companies and established companies have only one significant difference in the influences of covid 19 on the current competence need in the team/company (p=0.033)

Companies which have less than 10 years, and more than 10 years of operation show significant differences in:
\begin{itemize}
    \item the impact of WFH on leadership style (p=.024)
    \item current competence need in the team/ company (p=.003)
    \item the way business is done in your company’s sector over the next 5 years (p=0.018)
\end{itemize}

Companies located in different continents have the significant differences in:
\begin{itemize}
    \item the impact of WFH on maintaining organizational culture (p=0.015)
    \item the impact of WFH on leadership style (p=0.003)
    \item current competence need in the team/ company (p$<$0.001)
    \item the perception on how some Software Engineering Roles are affected, for instance Business Analyst/ Requirement Engineer (p=0.001), UX Designer (p=0.047), Project manager( p=0.023), Software Architect (p=0.017), Software Developer (p$<$0.001) and Scrum Master (p$<$0.001)
\end{itemize}

\subsubsection{Innovation activities}
\begin{table}[H]
    \centering
    {\small
    \begin{tabular}{|p{4cm}|c|c|c|c|}
    \hline
      \textbf{Questions} & \textit{Startups VS.} & \textit{Small vs.}  &  \textit{Recent vs.} & \textit{Among}\\
      & \textit{Established} & \textit{Large} & \textit{Long-time} & \textit{Continents}\\
      & & \textit{Team} & \textit{Companies} & \\
 \hline
 \hline
 changed commitment to innovation. & 0.520	&0.005**	&0.757	&0.317\\
 \hline
 increased investment on infrastructure. & 0.499	&0.016*	&0.149	&0.432\\
 \hline
 increased investment on internal R\&D. & 0.555	&0.012*	&0.696	&0.556 \\
\hline
 increased role of IP. & 0.593	&0.014*	&0.459	&0.163 \\
 \hline
 forced change in business model & 0.631	&0.561	&0.147	&0.774 \\
 \hline 
 fundamental change in near future & 0.434	&0.151	&0.018*	&0.051\\
 \hline
    \end{tabular}}
        \caption{\textbf{Perceived changes to Innovation activities across categories}}
    \label{tab:comp_inno}
\end{table}
Table \ref{tab:comp_inno} reveals that team size is the most useful factor to differentiate the perception of impact of WFH on innovation activities among our respondents. There is mostly no difference whether a company is startup or established about innovation activities.

\subsubsection{Resilience factors}
 
Table \ref{tab:comp resi} reveals that team size is the most useful factor to differentiate the perception of resilience factors among our respondents. There is no difference whether a company is startup or established when considering resilience factors.

\begin{table}[H]
    \centering
   
    {\small
    \begin{tabular}{|c|c|c|c|c|}
    \hline
      \textbf{Variables} & \textit{Small VS.} & \textit{Small VS.}  &  \textit{Recent VS.} & \textit{Among}\\
      & \textit{Established} & \textit{Large} & \textit{Long-time} & \textit{Continents}\\
      & & \textit{Team} & \textit{Companies} & \\
 \hline
 \hline
 Team & 0.114 &0.033* &0.385 &0.875 \\
 \hline
 Leadership & 0.217	& $<$ .001**	& 0.179&	0.013*\\
 \hline
 Financial Capital & 0.444 &	0.045* &	0.022* &	0.372 \\
\hline
 Social Capital & 0.740	&0.013*	&0.644	&0.228 \\
 \hline
 Process \& & 0.925	&0.114	&0.196	&0.110\\
 Working Practice &&&&\\
 \hline 
 Tools \& & 0.194 &	0.016*	&0.054	&0.734\\
 Infrastructures &&&&\\
 \hline 
 Agility & 0.119&	0.012*	&0.565	&0.060\\
 \hline
    \end{tabular}}
     \caption{\textbf{Difference among groups in perceived resilience factors}}
    \label{tab:comp resi}
\end{table} 

\begin{tcolorbox}
PO9: There is no difference found in terms of perception of the impact of WFH to project management, resilience and innovation activities between startups and established companies. However, there is a statistical difference between respondents who are in a small team and those is in a large team.
\end{tcolorbox}
\section{Discussion}

The summary of our findings is given in Table \ref{tab:summaryfindings}. In this section we discussed how our findings relate to existing studies, threats to validity and implications for research and practice.
\begin{table}[H]
\centering
\begin{tabular}{|p{3cm}|p{8.8cm}|p{1.1cm}|}
\hline
\textbf{RQs} & \textbf{Research Findings} & \textbf{POs}\\ \hline
RQ1 - How does the  of working in a software project change when WFH is adopted?    & Collaboration across different cities and across different time-zones tends to increase. Collaboration across the same city, and across different time zones tends to be the same. There is a wide range of collaboration and communication tools available and used among WFH software engineers. Various new coordination and control mechanisms are introduced and implemented for WFH, most of them based on formal procedures. & PO1, PO2, PO3\\ \hline
RQ2 - How is software project management impacted when WFH is adopted?   & There is a mixed perception on the impact of WFH on project planning and control. The majority tend to experience certain delays due to the new working environment shift. The majority tends to perceive a negative impact of the shift to new working environment on maintaining organizational culture. Project managers and business analysts seem to be the most impacted roles during the shift to WFH, Software developers seem to be least impacted. & PO4, PO5, PO6 \\ \hline
RQ3 - How are innovation activities in software companies impacted by the WFH situation?     & It seems that at the individual level, the majority experienced a decrease in motivation for innovation. However, at the organizational level, many agreed that their companies invested more on innovation infrastructure. & PO7\\ \hline
RQ4 - How is resilience achieved by software companies under WFH? & Team, Agility and Leadership are the most important perceived factors for software companies to achieve organizational resilience. & PO8\\ \hline 
RQ5 - Is there any difference between startups and established companies regarding the above impact? & There is no difference found in terms of perception on the impact of WFH to project management, resilience, and innovation activities between startups and established companies. However, there is a statistical difference between respondents who are in a small team and those in a large team. & PO9 \\ \hline
\end{tabular}
\caption{\textbf{Summary of the findings}}
\label{tab:summaryfindings}
\end{table}

\subsection{Relations to existing studies}

Ford et al. conducted both qualitative and quantitative research on software engineers at Microsoft during Spring 2020. \cite{ford_tale_2021}. The authors reported a mixed experience for software engineers when working from home, and for the same factor, one engineer can perceive a benefit and another can perceive it as a challenge. This might be an explanation for the overall mixed picture of positive and negative impact of WFH on software projects from our dataset.

Gregory et al. discussed challenges for requirement engineers when WFH practices are enforced in a company \cite{gregory_requirements_2021}. Our findings provide a comparative view among software engineering roles in projects, and highlight the perceived challenges especially with the role of project managers and requirements engineers.

Smite et al. revealed that during WFH context, software engineers continue committing code and carry out their daily duties without significant disruptions \cite{smite_changes_2022}. This also aligns with findings that developers are found least impacted by the new situation. Smite et al. and Rodeghero et al. described socializing, communication and collaboration and onboarding practices that have been adapted to WFH context \cite{smite_changes_2022,rodeghero_please_2021}. While we cover some of these practices in our survey, i.e. turn-on camera, social network for socializing, more meetings, etc. our study show the statistics on how popular these practices are among our respondents.

Our study shows that while individual motivation tends to decrease during the WFH context, the investment of companies in innovation infrastructure and resources tended to increase during this time. This might be explained by the fact that companies new to a WFH setting may have lacked proper digital infrastructure, therefore needing to increase investment in that area. \cite{bai_digital_2020}.

Kettunen et al. conducted a survey on Agile practices during the pandemic in Sweden and Finland \cite{kettunen_impacts_2021}. When asked about the impact of the pandemic impact their companies, the authors reported that 53\% of the respondents had perceived negative impact and 33\% of the respondent had perceived positive impacts. Our tendency aligns with this observation, however, we provide specific details on which software engineering roles are affected and which aspect of companies are perceived either negatively or positively. We also agree with Kettunen, in finding that Agility is one of the top three factors for software companies to remain resilient during the pandemic. We have added  evidence on teh strength of the role of agility \cite{sonjit_homeworking_2021,aldianto_toward_2021} and leadership \cite{aldianto_toward_2021} in maintaining software companies' resilience and performance, both important  in a startup context.

\subsection{Threats to validity}

We will discuss validity according to the four perspectives presented by Wohlin et al. \cite{wohlin_experimentation_2012}, complemented by survey-specific validity aspects \cite{fink_how_2022,molleri_survey_2016}.

Construct validity is concerned with the relationship between a theory behind an investigation and its observation \cite{wohlin_experimentation_2012}. As the goal of the survey is to gain industrial insights on WFH and relating practices, we do not aim at fully developing or validating hypotheses. However, the observations from our study can give hints for further research about working environment (i.e. WFH) and different properties of software companies and software startups. To enhance construct validity, we used validated scales for software engineering activities and project management. We are also confident about the confidentiality and anonymity of the respondents, hence reducing as much bias as possible.

Internal validity deals with the relationship between a treatment and its results \cite{wohlin_experimentation_2012}. We have a filtering question so that only respondents who experienced an impact on their work and their companies can answer questions.

An inherent threat to survey research is that surveys can only reflect respondents’
perceptions rather than objective measurement. To make questions understood in the same ways by all respondents, we reviewed and revised several times (Section 4.2). The survey versions were reviewed by people from representative countries to reduce the possible misunderstanding due to language or cultural differences.

Our findings are based on a reasonable but still limited number of respondents. With 45 questions (including complex questions and open text questions), our survey is among long questionnaires that are conducted in software engineering. It takes in average 18 minutes 50 seconds to complete the survey. Hence, 297 valid data points is a good number for the scope of the survey. This is comparable with previous survey in software engineering, as seen in Table \ref{tab:surveyinseliterature}. 

\begin{table}[H]

\begin{tabular}{|p{1.2cm}|p{2.5cm}|p{1.2cm}|p{5cm}|p{1.8cm}|}
\hline
\textbf{\# questions }& \textbf{\# valid responses} &\textbf{Publ. Year} & \textbf{Publ. venues}                        & \textbf{Ref.}  \\ \hline
35               & 228 organizations      & 2017 & Empirical Software Engineering           &   \cite{fernandez_naming_2017}         \\ \hline
46               & 202 software engineers & 2015 & Information and Software Technology      &   \cite{garousi_survey_2015}         \\ \hline
23               & 101 engineers          & 2015 & IEEE Transaction in Software Engineering &   \cite{de_la_vara_industrial_2016}         \\ \hline
25               & 69 practitioners       & 2019 & IEEE Software                            &    \cite{kuhrmann_hybrid_2019}        \\ \hline
45               & 297 companies          & 2021 &                                          & this study \\ \hline
\end{tabular}
\caption{\textbf{Existing industrial surveys in software engineering literature}}
\label{tab:surveyinseliterature}
\end{table}

External validity is concerned with the generalization of the conclusions \cite{wohlin_experimentation_2012}. We cannot make a generalized conclusion from our study. Proper sampling is very difficult to conduct due to no credible sampling frames (population lists) for the units of analysis in software engineering researchers study \cite{amir_poster_2018, ralph_pandemic_2020}. Our unit of analysis is the software company, but we are not able to estimate representativeness of our population due to the unavailability of empirical data from each industry sector and each country. A different result might be observed with a different sample. However, the survey can be repeatedly conducted and new results can be synthesized with what is reported in this study. We note that it is seen as uncommon to have a survey on a narrow topic in software engineering with more than 100 valid responses \cite{de_la_vara_industrial_2016}.

Conclusion validity is concerned with obstacles to draw correct conclusions from a study \cite{wohlin_experimentation_2012}  . Although we did not conduct random sampling, we have tried our best to diversify the respondents regarding their geographical locations (from 35 countries), industrial sectors (more than 18 sectors), company types (startups and established companies) and team size. Conclusion validity is further strengthen by data triangulation, having consistent observations from both quantitative and qualitative data.

%\subsection{Implications to research and practice}

\section{Conclusions}

The pandemic has caused a major disruption worldwide. Research started to reveal what has happened to the software industry, i.e the development of software product and establishment of software startups. One thing is clear to many, at the time of writing in December 2021, a return to previous business and working condition is unlikely, and it is predicted that the future of software industry will based on Working from home in some way. This study reports a results from a global-scale, cross-sectional survey on WFH and its impact on software project, innovation, and resilience. We collected answers to 46 questions from 297 respondents around the world.

Our result agrees with recent studies on the topic, for instance, the mixed perceived impact of WFH at individual and team levels, the new coordination and control mechanisms, important factors for organisational resilience. We highlighted the difference in perception from different software engineering roles, and between small and large teams. We also highlighted the role of team, agility and leadership for achieving resilience in software companies.

In the near future the working environment will have a significant impact in where, when and how WFH is adopted in software projects, hence it is important to understand its impact on organizational and team factors besides productivity or well-beingness. Our study suggests that many current challenges of WFH can be cured by looking into team, leadership and management and coordination mechanisms. We believe that companies with a cohesive and agile team with a suitable leadership approach and appropriate investment on team coordination and support are likely to stay resilient, innovative and grow post-pandemic. There might not be a one-size-fit all approach for an optimal WFH configuration, but at least we can see the effect of team size and geographical locations on the perceived performance of WFH team. Alternatively, national, organizational, and cultural dimensions should be taken into account when defining a WFH strategy. 

As future work, we plan to conduct more analysis, i.e. factor analysis or regression analysis to understand the possible relationship between WFH practices, context factors and productivity. We aim at conducting the analysis at both individual, team and organizational level for comparison purpose. In future work, we also plan to explore the qualitative data to explain what is observed from the survey.

\bibliographystyle{elsarticle-num}
\bibliography{ref}

\begin{thebibliography}{10}
\bibitem{blank_steve_2010}
S.~Blank, Steve blank an {MVP} is not a cheaper product, it’s about smart
  learning.
\end{thebibliography}

%\printbibliography





\end{document}